\documentclass[epsfig,usenatbib]{mnras}
\usepackage{epsfig}
\usepackage{natbib}
\usepackage{amsmath}

\def\gtsima{$\; \buildrel > \over \sim \;$}
\def\ltsima{$\; \buildrel < \over \sim \;$}
\def\gtrsim{\lower.5ex\hbox{\gtsima}}
\def\lesssim{\lower.5ex\hbox{\ltsima}}

\begin{document}
\title[The host galaxies of double compact objects]{The host galaxies of double compact objects merging in the local Universe}
\author[Mapelli et al.]
       {Michela Mapelli$^{1,2,3}$, Nicola Giacobbo$^{1,2,3,4}$, Mattia Toffano$^{4}$, Emanuele Ripamonti$^{4}$,
         \newauthor Alessandro Bressan$^{5}$, Mario Spera$^{1,3}$, Marica Branchesi$^{6}$ 
\\
$^1$Institut f\"ur Astro- und Teilchenphysik, Universit\"at Innsbruck, Technikerstrasse 25/8, A--6020, Innsbruck, Austria\\
$^2$INAF-Osservatorio Astronomico di Padova, Vicolo dell'Osservatorio 5, I--35122, Padova, Italy, {\tt michela.mapelli@oapd.inaf.it}\\
$^3$INFN, Milano Bicocca, Piazza della Scienza 3, I--20126 Milano, Italy\\
$^4$Physics and Astronomy Department Galileo Galilei, University of Padova, Vicolo dell'Osservatorio 3, I--35122, Padova, Italy\\
$^5$Scuola Internazionale Superiore di Studi Avanzati (SISSA), Via Bonomea 265, I-34136, Trieste, Italy\\
}
\maketitle \vspace {7cm }
\bibliographystyle{mnras}
 
\begin{abstract}
We investigate the host galaxies of compact objects merging in the local Universe, by combining the results of binary population-synthesis simulations with the Illustris cosmological box. Double neutron stars (DNSs) merging in the local Universe tend to form in massive galaxies (with stellar mass $>10^{9}$ M$_\odot$) and to merge in  the same galaxy where they formed, with a short delay time between the formation of the progenitor stars and the DNS merger. In contrast, double black holes (DBHs) and black hole -- neutron star binaries (BHNSs) form preferentially in small galaxies (with stellar mass $<10^{10}$ M$_\odot$) and merge either in small or in larger galaxies, with a long delay time. This result is an effect of metallicity: merging DBHs and BHNSs form preferentially from metal-poor progenitors ($Z\leq{}0.1$ Z$_\odot$), which are more common in high-redshift galaxies and in local dwarf galaxies, whereas merging DNSs are only mildly sensitive to progenitor's metallicity and thus are more abundant in massive galaxies nowadays. The mass range of DNS hosts we predict in this work is consistent with the mass range of short gamma-ray burst hosts. 
\end{abstract}

%

\section{Introduction}
On September 14 2015, the two LIGO interferometers \citep{LIGOdetector}  obtained the first direct detection of gravitational waves (GWs). During the second observing run, the Virgo detector \citep{VIRGOdetector} joined the network, drastically improving the sky localization of GW events \citep{abbottGW170814,abbottGW170817}. Six confirmed GW events have been reported so far, five of them associated to merging double black holes (DBHs \citealt{abbottGW150914,abbottastrophysics,abbottGW151226,abbottO1,abbottGW170104,abbottGW170608,abbottGW170814}) and the remaining one, GW170817, interpreted as a double neutron star (DNS) merger \citep{abbottGW170817}.

No electromagnetic counterpart was identified for the five DBH events, whose host galaxies remain unknown. In contrast, GW170817 was associated to electromagnetic emission spanning almost the entire electromagnetic spectrum, from gamma rays to radio wavelengths \citep{abbottmultimessenger,abbottGRB,goldstein2017,savchenko2017,margutti2017,coulter2017,soares-santos2017,chornock2017,cowperthwaite2017,nicholl2017,pian2017,alexander2017}. The detection of the electromagnetic counterpart led to the indisputable identification of the host galaxy as NGC~4993 \citep{coulter2017,soares-santos2017}, an early-type galaxy with stellar mass $0.3-1.2\times{}10^{11}$ M$_\odot$ \citep{im2017}, redshift $z\sim{}0.009783$ \citep{levan2017}, and mostly old (but not exclusively old) stellar population \citep{levan2017,fong2017}.



Several tens of new GW detections are expected in the forthcoming observing runs of LIGO and Virgo, possibly associated with the identification of the host galaxy. In preparation for future detections, it is crucial to study the properties of the host galaxies of merging DNSs, DBHs and black hole -- neutron star binaries (hereafter BHNSs) with theoretical models, for a number of reasons. First, the comparison of different models against future detections is a vital test for models of compact-object binary formation, which are currently affected by a plethora of uncertainties (see e.g. \citealt{tutukov1973,flannery1975, bethe1998, portegieszwart1998,  portegieszwart2000,belczynski2002,voss2003, podsiadlowski2004,podsiadlowski2005,tauris2006,belczynski2007,  bogomazov2007,  dominik2012, dominik2013,  dominik2015,  mennekens2014, spera2015,tauris2015, tauris2017, demink2015,demink2016,marchant2016,chruslinska2018,mapelli2017,giacobbo2018,kruckow2018,shao2018}).


Second, studying the properties of host galaxies from simulations can provide us with astrophysically motivated criteria to localize the host galaxy of a GW event, even if the electromagnetic counterpart is not observed. Namely, if  more than one galaxy is found in the error box of GW detectors, astrophysically motivated criteria could help us telling which is the most likely host galaxy.


Several papers have attempted  studying the properties of DBH host galaxies, with mixed results (e.g. \citealt{lamberts2016,dvorkin2016,oshaughnessy2017,mapelli2017,schneider2017,elbert2017,cao2018,mapelli2018}). \cite{lamberts2016}, \cite{mapelli2017} and \cite{belczynski2016}  reconstruct the environment of GW150914 by accounting for the evolution of the star formation rate density and of stellar metallicity across cosmic time. All these studies conclude that GW150914 progenitors could have formed either at high redshift ($\gtrsim{}2$) or at low redshift ($\sim{}0.2$).

\cite{lamberts2016} find that low-redshift progenitors are metal poor and form mostly in dwarf galaxies, whereas high-redshift progenitors are mostly metal rich ($Z\gtrsim{}0.1$ Z$_\odot{}$) and form in large galaxies. In contrast, \cite{belczynski2016} and \cite{mapelli2017} (but see also \cite{schneider2017,spera2017,giacobbo2018,giacobbo2018b}) find that most progenitors of GW150914-like systems have metallicity $Z\leq{}0.1$ Z$_\odot$. In particular, \cite{schneider2017} find that most GW150914 progenitors form in dwarf galaxies with stellar mass $<5\times{}10^6$ M$_\odot$, but then merge when they are hosted by large ($>10^{10}$ M$_\odot$) star forming galaxies.

Considering not only GW150914-like systems but all DBHs, \cite{elbert2017} suggest that DBH mergers are mostly localized in dwarf galaxies if the merger time-scale is short, but in massive galaxies otherwise. By using zoom-in cosmological simulations, \cite{oshaughnessy2017} suggest that dwarf galaxies overabundantly produce DBH mergers. Finally \cite{cao2018} investigate the host galaxies of DBHs by applying a semi-analytic model to the Millennium-II N-body simulation \citep{boylan2009}. They find that DBHs merging at redshift $z\lesssim{}0.3$ are located mostly in massive galaxies (stellar mass $\gtrsim{}2\times{}10^{10}$ M$_\odot$).

In comparison to DBHs, few studies have focused on the environment of DNS mergers \citep{perna2002,dominik2013,giacobbo2018b,mapelli2018}. The general understanding is that DNSs are less affected by progenitor's metallicity than DBHs.

In this paper, we combine information from catalogues of merging compact objects with the snapshots of the Illustris-1 cosmological simulation \citep{vogelsberger2014a,vogelsberger2014b}, to obtain information on the host galaxies of DNSs, BHNSs and DBHs merging in the local Universe. The catalogues of merging compact objects are obtained with our new population-synthesis code {\sc MOBSE} \citep{giacobbo2018}, which includes up-to-date models for stellar winds and different flavours of supernovae (SNe).

\section{Methods}\label{sec:methods}
        {\sc MOBSE} is an upgrade of the population-synthesis code {\sc BSE} \citep{hurley2000,hurley2002}, to include recent models of stellar winds \citep{vink2001,vink2005,graefener2008,vink2011}, core-collapse SNe \citep{fryer2012}, electron-capture SNe \citep{giacobbo2018c}, pair-instability and pulsational pair-instability SNe \citep{woosley2017,spera2017}.

        Mass loss by stellar winds is implemented in {\sc MOBSE} as $\dot{M}\propto{}Z^\beta$, where $\beta{}=0.85$ for Eddington ratio $\Gamma\le{}2/3$, $\beta{}=2.45-2.4\,{}\Gamma{}$ for $2/3\leq{}\Gamma{}\leq{}1$ and $\beta{}=0.05$ for $\Gamma{}>1$ \citep{chen2015}. This accounts for the dependence of stellar winds on both metallicity and Eddington ratio. The combination of this prescription for stellar winds with the adopted SN models produces a distribution of black hole (BH) masses which strongly depends on metallicity: at low metallicity ($Z\lesssim{}0.0002$), BHs can reach masses of $\sim{}60$ M$_\odot$, while at solar metallicity the maximum BH mass is $\sim{}25$ M$_\odot$ \citep{giacobbo2018}. The statistics of merging BHs, neutron stars (NSs) and BHNSs obtained with {\sc MOBSE} is nicely consistent with the masses and rates inferred from the LIGO-Virgo collaboration \citep{abbottO1,abbottGW170817}. For more details about {\sc MOBSE}, we refer to \cite{giacobbo2018} and \cite{giacobbo2018c}. 

For this work, we use the catalogue of merging compact objects obtained from run CC15$\alpha{}$5 of \cite{giacobbo2018b}, because this simulation best matches the cosmic merger rate density inferred from LIGO-Virgo results \citep{mapelli2018}. In run~CC15$\alpha{}$5 we have assumed high efficiency of common-envelope ejection (described by the parameter $\alpha=5$) and low SN kicks (described by a Maxwellian curve with one-dimensional root mean square $\sigma{}=15$ km s$^{-1}$). Simulation CC15$\alpha{}$5 consists of 12 sub-sets corresponding to metallicity $Z/{\rm Z}_\odot{}=0.01$,   0.02,   0.04,   0.06,   0.08,   0.1,   0.2,   0.3,   0.4,   0.6,   0.8, and 1.0 (here we assume $Z_\odot{}=0.02$). In each sub-set  we  have simulated $10^7$ stellar binaries, for a total number of $1.2\times{}10^8$ simulated binaries. Considering that the formation of compact object binaries has a similar trend with metallicity also in the other simulations presented in \cite{giacobbo2018b}, we  expect that our results do not depend significantly on the choice of this run. 

The catalogue of merging compact objects derived from population-synthesis simulations is combined with the cosmological simulation
as described in \cite{mapelli2017} and \cite{mapelli2018}. In particular, we randomly associate a number of merging compact-object binaries to each Illustris-1 stellar particle based on its initial mass, formation redshift and metallicity. The main difference with respect to \cite{mapelli2017} and \cite{mapelli2018} is that in the current paper we do not use the metallicity of the Illustris-1 particles, but we use the metallicity derived from the empirical mass -- metallicity relation, as we describe in the following.

The Illustris-1 is the highest resolution hydrodynamical simulation run in the frame of the Illustris project \citep{vogelsberger2014a,vogelsberger2014b,nelson2015}. It covers a comoving volume of $(106.5\,{}{\rm Mpc})^3$, and has an initial dark matter and baryonic matter mass resolution of $6.26\times{}10^6$ and $1.26\times{}10^6$ M$_\odot$, respectively \citep{vogelsberger2014a,vogelsberger2014b}. The size of the Illustris-1' box ensures that we include the most massive haloes, while dwarf galaxies are  unresolved. 

 The model of sub-grid physics adopted in the Illustris-1 is known to produce a mass-metallicity relation \citep{vogelsberger2013,genel2014,genel2016} which is sensibly steeper than the observed one (see the discussion in \citealt{vogelsberger2013} and \citealt{torrey2014}). Moreover, the simulated mass-metallicity relation does not show the observed turnover at high stellar mass ($\gtrsim{}10^{10}$ M$_\odot{}$).

\begin{table}
\begin{center}
\caption{\label{tab:tableB2}
Best fit parameters for the mass-metallicity relation in equation~\ref{eq:maiolino} at different redshift. The values of $M_0$ and $K_0$ at $z=3.5$ come from Mannucci et al. (2009), while the other values come from Maiolino et al. (2008).}
 \leavevmode
\begin{tabular}[!h]{ccc}
\hline
$z$ &  $\log{M_0}$ & $K_0$\\ 
\hline
0.07  & 11.18 & 9.04 \\
0.7   & 11.57 & 9.04 \\
2.2   & 12.38 & 8.99 \\
3.5  &  12.28  & 8.69    \\
\noalign{\vspace{0.1cm}}
\hline
\end{tabular}
\begin{flushleft}
\footnotesize{Column~1: redshift; column~2 and 3: values of the parameters in equation~\ref{eq:maiolino} at different redshift.}
\end{flushleft}
\end{center}
\end{table}

For these reasons in this paper we  override the metallicity of a given Illustris-1 star particle with the metallicity we expect from the observed mass-metallicity relation. For the observed relation, we adopt the fitting formula by \cite{maiolino2008} and \cite{mannucci2009}:
\begin{equation}\label{eq:maiolino}
12+\log{({\rm O}/{\rm H})}=-0.0864\,{}(\log{M_\ast{}} - \log{M_0})^2\,{}+K_0,
\end{equation} 
where $M_\ast$ is the total stellar mass of the host galaxy in solar masses, while $M_0$ and $K_0$ are given in Table~\ref{tab:tableB2}. For intermediate redshifts between those in Table~\ref{tab:tableB2}, we obtain the metallicity by linear interpolation. At redshift $z<0.07$ ($z>3.5$) we simply use the same coefficients as for $z=0.07$ ($z=3.5$). The new metallicity of each Illustris-1' star is randomly extracted from a Gaussian distribution with mean value given by equation~\ref{eq:maiolino} (where $M_\ast{}$ is the total stellar mass of the sub-halo hosting the Illustris-1' star) and standard deviation $\sigma{}=0.3$ dex (accounting for metallicity dispersion within galaxies). We also repeated our calculations for a larger (smaller) scatter $\sigma{}=0.5$ ($\sigma{}=0.2$) \citep{mapelli2017} and we checked that our main results are not significantly affected by this assumption.

Following the aforementioned procedure, we can directly obtain information on the stellar mass of the galaxy where a compact object binary merges (hereafter $M_{\rm merg}$) or where the stellar progenitors of the compact object binary form (hereafter $M_{\rm form}$). In this paper, we consider only compact objects merging at $z\leq{}0.024$, because we are interested in the host galaxies of mergers happening in the very nearby Universe. NGC~4993, the host galaxy of GW170817, falls within this redshift bin, because its redshift is $z=0.009783$ \citep{levan2017}. 
Actually, $z=0.024$ is the redshift corresponding to the third last snapshot of the Illustris simulation. 
In future studies, we will investigate how the properties of the host galaxies change as a function of redshift, considering also mergers occurring at higher redshift. 

\section{Results}\label{sec:results} 

\begin{figure}
\center{{
\epsfig{figure=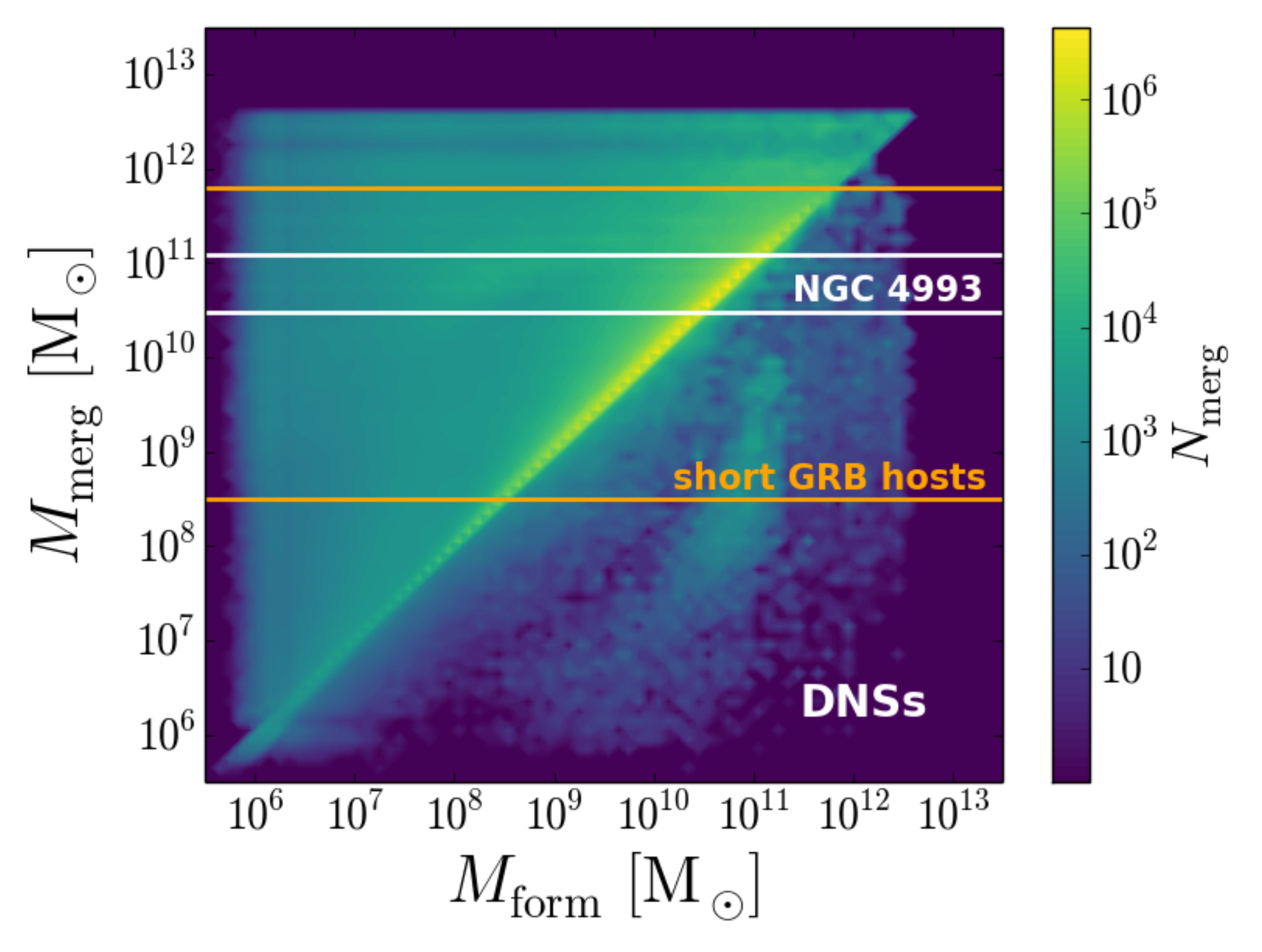,width=9cm} 
\epsfig{figure=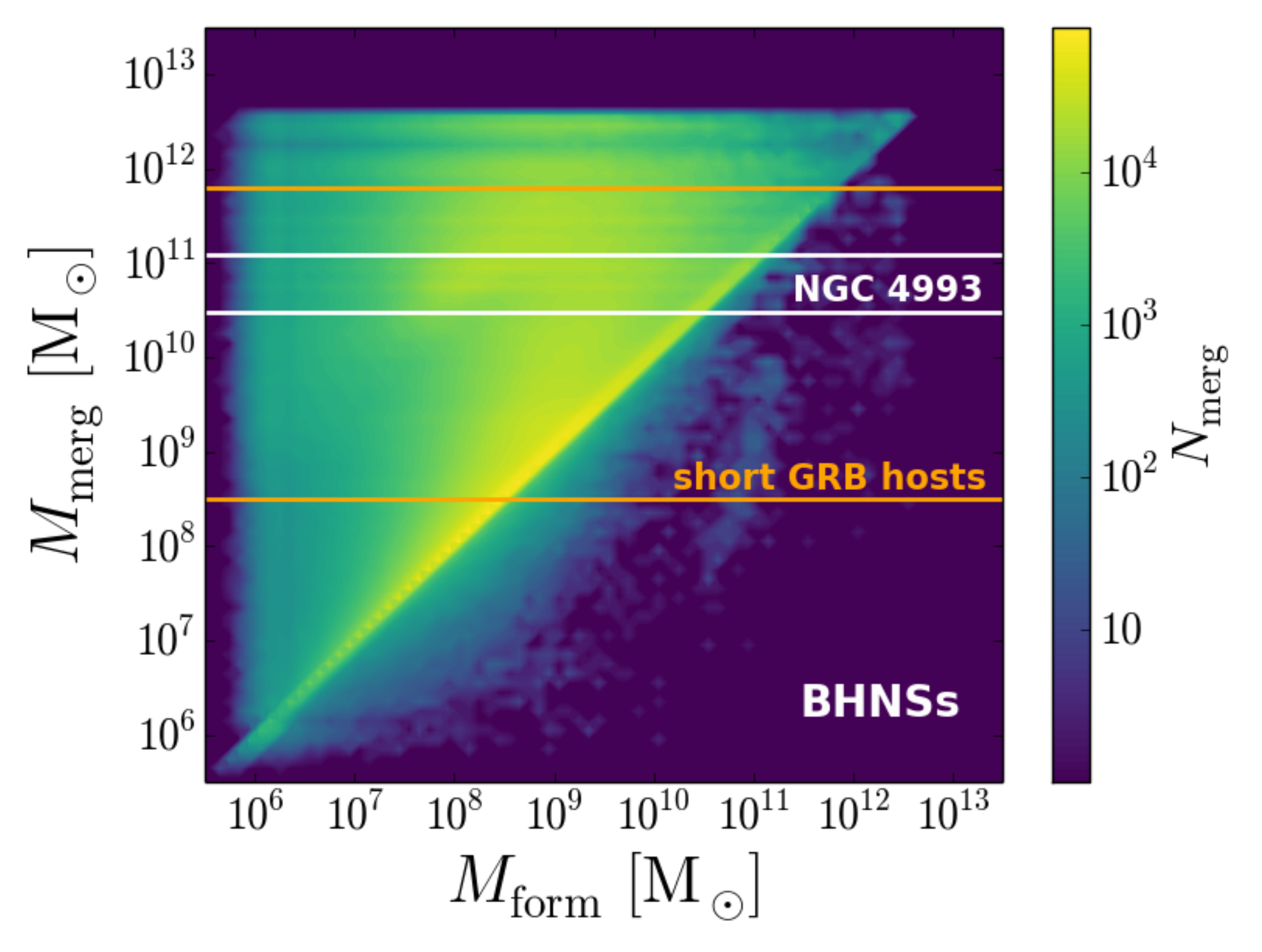,width=9cm} 
\epsfig{figure=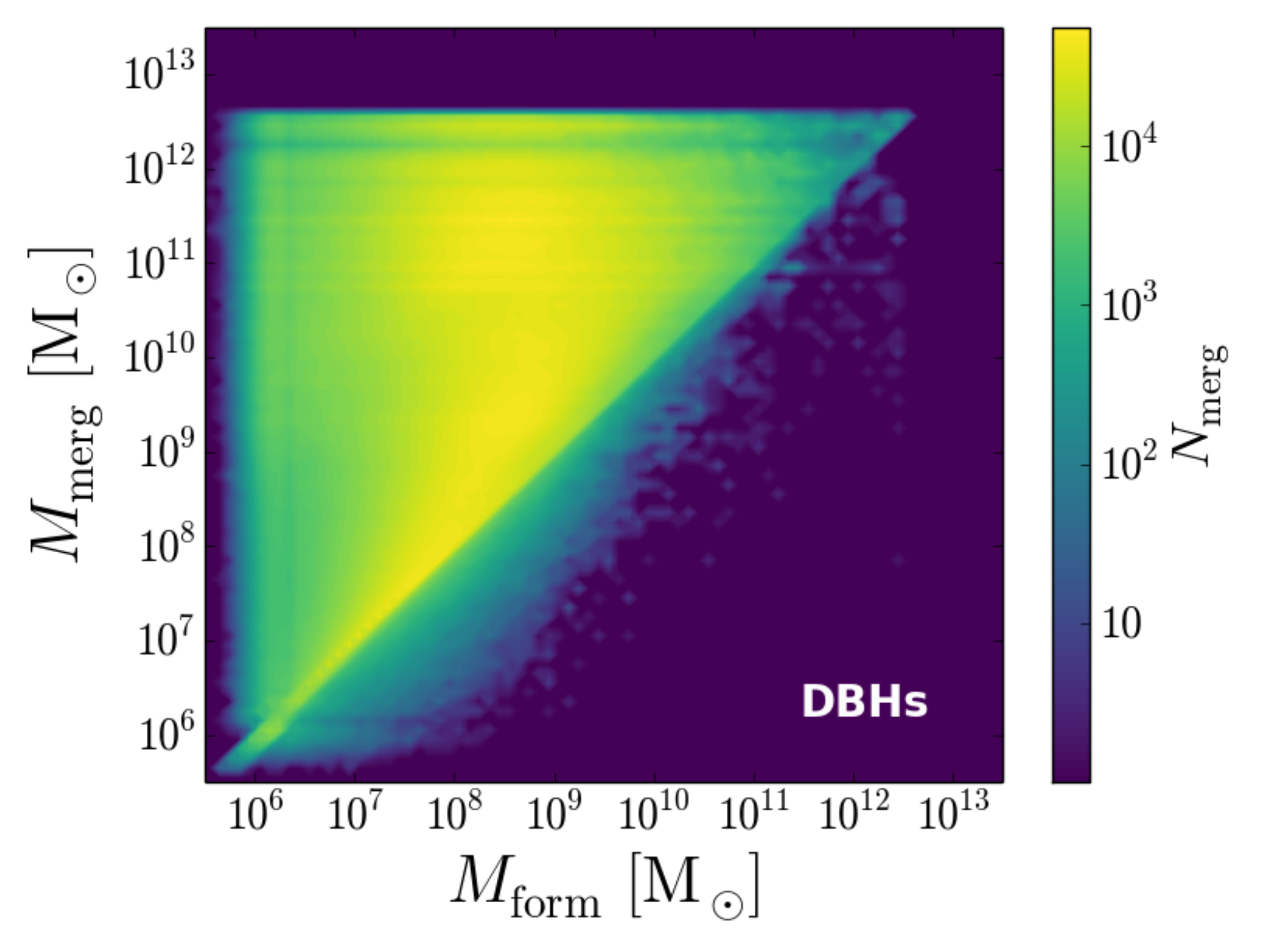,width=9cm} 
}}
\caption{\label{fig:figmergmform_contour}
Stellar mass of galaxies where compact-object binaries merge ($M_{\rm merg}$) versus stellar mass of galaxies where compact-object binaries form ($M_{\rm form}$). We show only compact-object binaries merging at $z\leq{}0.024$. The colour-coded map (in logarithmic scale) indicates the number of merging compact objects per cell. The cell size is $\log{\delta{}M_{\rm merg}/M_\odot}\times{}\log{\delta{}M_{\rm form}/M_\odot}=0.1\times{}0.1$. Top panel: merging DNSs; central panel: merging BHNSs; bottom panel: merging DBHs. The orange lines in the top and central panel show the extremes of the mass range of short gamma-ray burst hosts \citep{leibler2010}. The two white lines in the same panels show the uncertainty range of the stellar mass of NGC~4993, the host galaxy of GW170817 \citep{im2017}.}
\end{figure}
The top panel of Figure~\ref{fig:figmergmform_contour} shows the stellar mass of the galaxy where a DNS merges ($M_{\rm merg}$) versus the stellar mass of the galaxy where the stellar progenitor of the merging DNS formed ($M_{\rm form}$), if we consider only DNSs merging at redshift $z\leq{}0.024$. From this figure, it is apparent that a large fraction of DNSs merging in the local Universe form and merge in the same galaxy (the most densely populated region is the diagonal of the plot).

\begin{figure}
\center{{
    \epsfig{figure=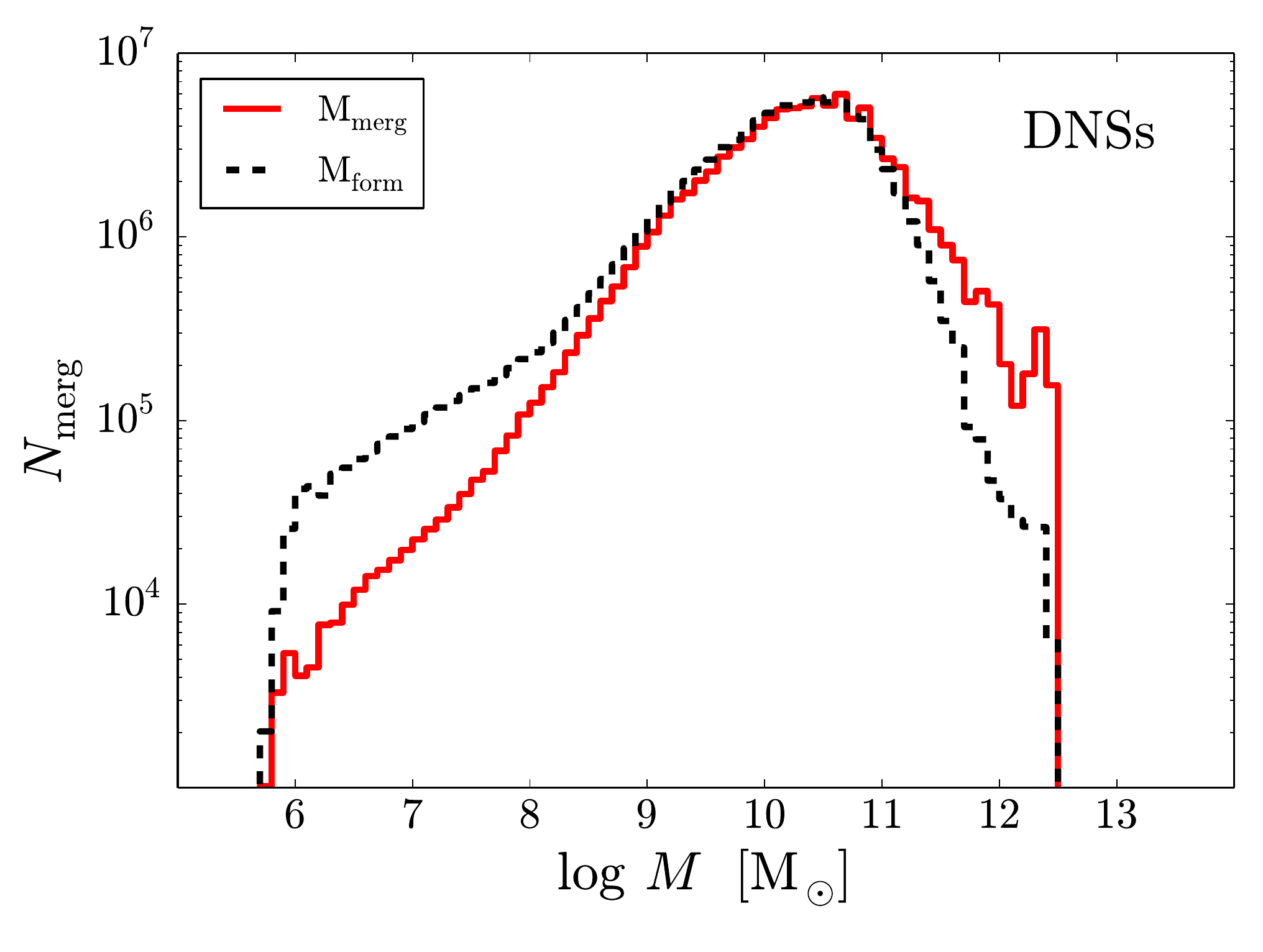,width=7cm} 
    \epsfig{figure=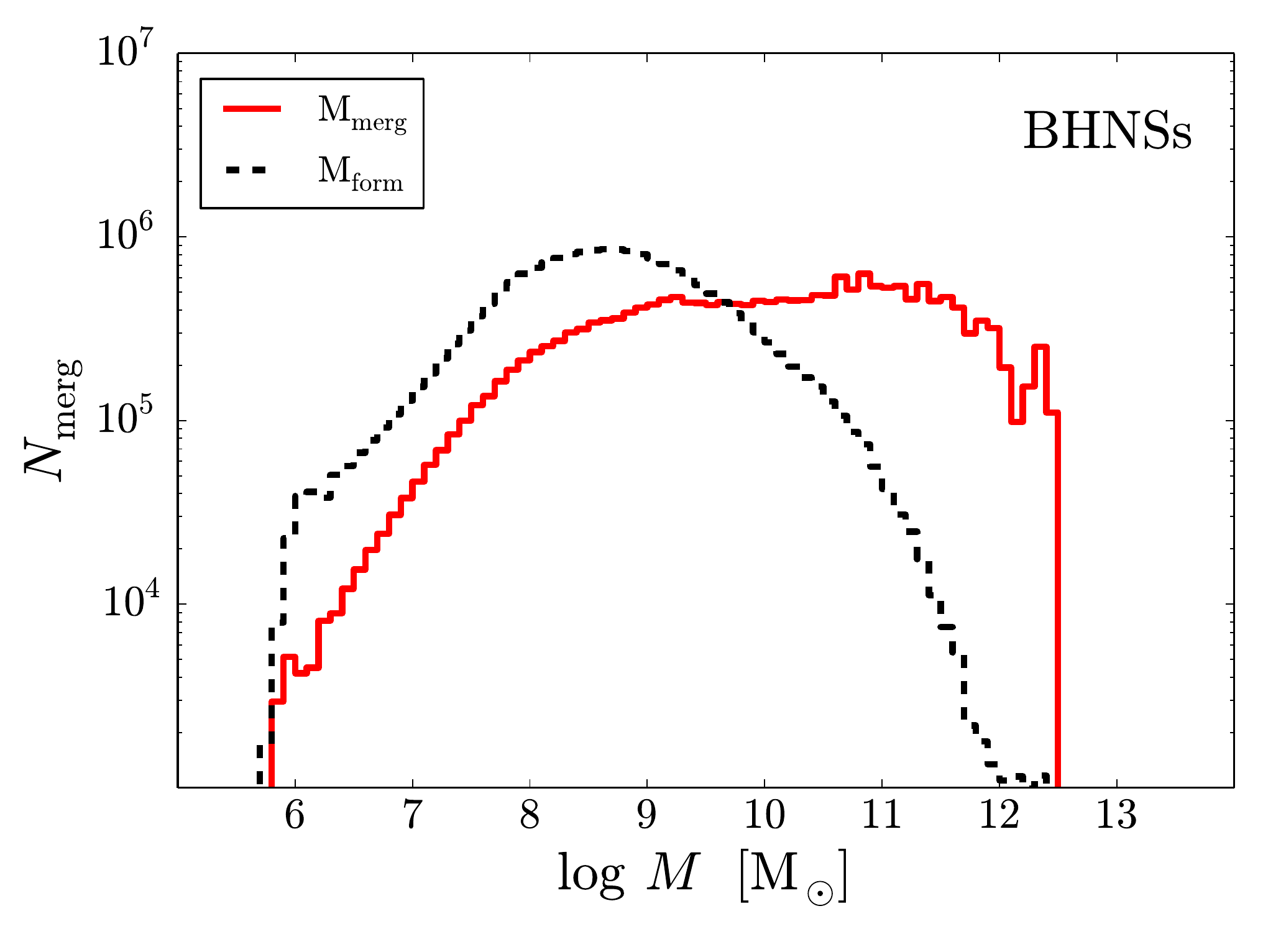,width=7cm} 
    \epsfig{figure=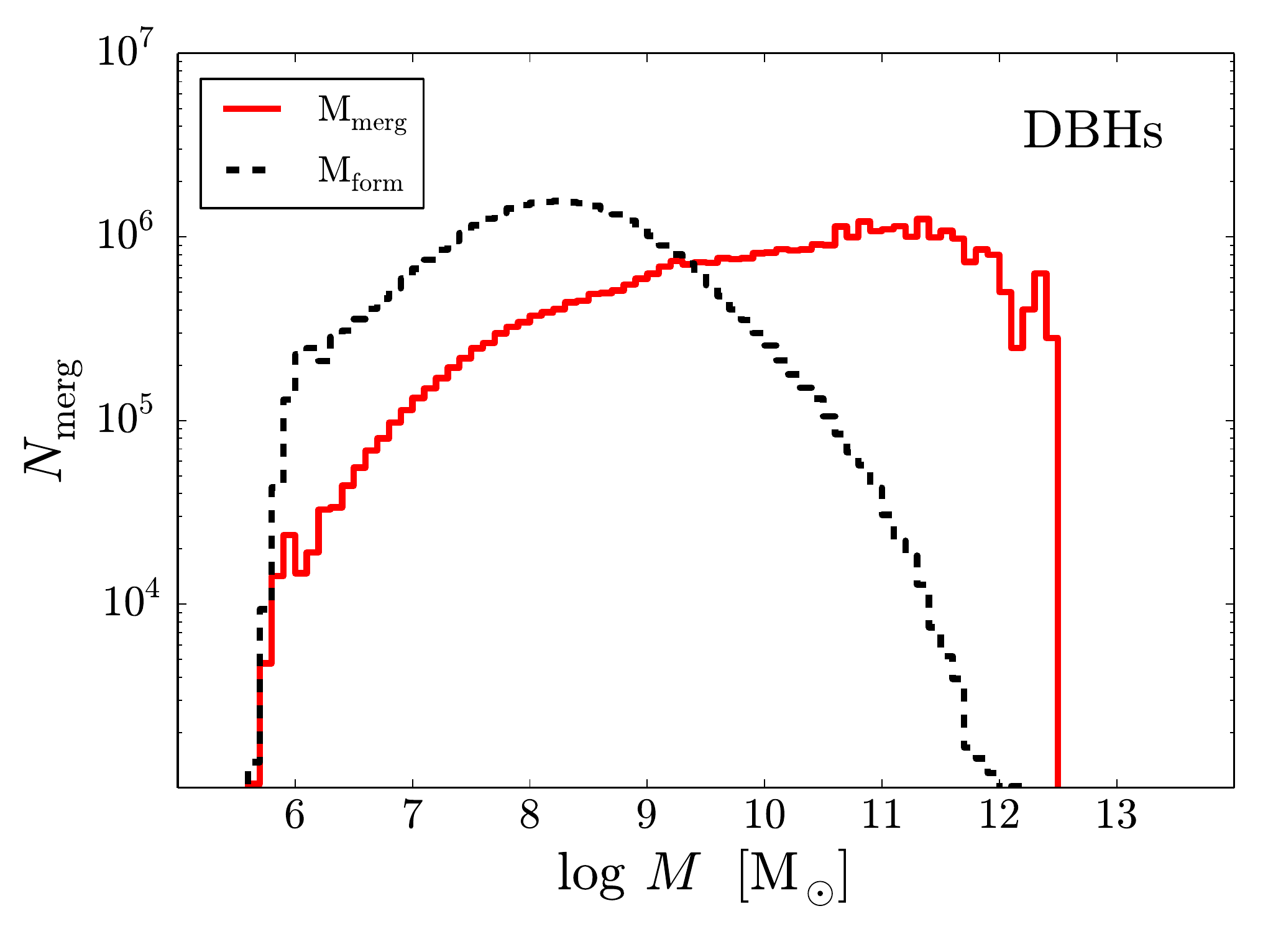,width=7cm} 
}}
\caption{\label{fig:figmergmform}
Distribution of the total masses of the host galaxies of merging DNSs (top), BHNSs (middle) and DNSs (bottom panel). Red solid lines: $M_{\rm merg}$; black dashed lines: $M_{\rm form}$.}
\end{figure}

The galaxies where such DNSs form and merge are predominantly massive, with typical stellar masses ranging from $\sim{}10^9$ to $\sim{}10^{12}$ M$_\odot$ (see the top panel of Figure~\ref{fig:figmergmform}). This mass range is consistent with the typical masses of the host galaxies of short $\gamma{}$-ray bursts (GRBs, $10^{8.5}-10^{12}$ M$_\odot$, \citealt{leibler2010,berger2014,troja2016}). Moreover, the estimated stellar mass of NGC~4993, which is the host galaxy of GW170817, is $0.3-1.2\times{}10^{11}$ M$_\odot$ \citep{im2017}, very close to the median mass of the distribution in Fig.~\ref{fig:figmergmform}.

The central and bottom panels of Fig.~\ref{fig:figmergmform_contour} show $M_{\rm merg}$ versus $M_{\rm form}$ for BHNSs and DBHs merging at redshift $z\leq{}0.024$, respectively. The host galaxies of both BHNSs and DBHs show a trend that is very different from that of DNSs. In fact, their progenitors tend to form in small galaxies ($\lesssim{}10^{10}$ M$_\odot$) and merge either in small galaxies or in larger ones. The typical mass of galaxies where DBHs and BHNSs form is $M_{\rm form}\sim{}10^7-10^9$ M$_\odot$, while they merge mostly in galaxies with $M_{\rm merg}>10^9$ M$_\odot$ (Fig.~\ref{fig:figmergmform}).

Actually, the peak of the distribution of $M_{\rm form}$ might be even lower than $M_{\rm form}\sim{}10^8$ M$_\odot$, because dwarf galaxies with mass $<10^9$ M$_\odot$ are unresolved in the Illustris-1 simulation. The analysis of small box high-resolution simulations (like e.g. the {\sc{GAMESH}} simulation, \citealt{schneider2017}) has shown that dwarf galaxies give a significant contribution to the population of GW150914-like systems across cosmic time.

 From Figure~\ref{fig:figmergmform_contour} we see that there is a relatively small number of DNS, BHNS and DBH mergers for which $M_{\rm merg}<M_{\rm form}$. In most cases this is a spurious numerical effect connected with the ability of the sub-halo finder algorithm to attribute stellar particles to the correct galaxy. 
  In fact, this effect is stronger for lower values of $M_{\rm form}$ and $M_{\rm merg}$ than for larger ones.

The different trend of DNS host galaxies with respect to DBHs and BHNSs is mainly an effect of metallicity. As we discussed in previous papers \citep{ziosi2014,mapelli2016,giacobbo2018,giacobbo2018b}, DBHs and BHNSs form preferentially from metal-poor progenitors, while DNSs form from both metal-poor and metal-rich progenitors with nearly the same efficiency \citep{giacobbo2018b,giacobbo2018c}.

Metal-poor stars are preferentially located in low-redshift dwarf galaxies and in high-redshift galaxies. Thus, DBHs and BHNSs form preferentially in these two environments. For this reason, in the local Universe we expect to observe i) mergers of DBHs and BHNSs formed in local metal-poor dwarf galaxies and with short delay time (the delay time $t_{\rm delay}$ is defined as the time elapsed between the formation of the stellar progenitors and the merger of the compact-object binary), and ii) mergers of DBHs and BHNSs formed in high-redshift galaxies and with long delay times.

While DBHs and BHNSs formed in local metal-poor galaxies and with short delay time tend to merge in the same galaxy where they formed, DBHs and BHNSs formed in high-redshift galaxies and with long delay time tend to merge in galaxies larger than the ones where they formed, because their original hosts have merged into larger galaxies.

In contrast, merging DNSs form from metal-rich and metal-poor progenitors nearly with the same efficiency. Thus, we expect that merging DNSs in the local Universe are mostly in large galaxies, where most of the stellar mass is confined.

\begin{figure}
\center{{
\epsfig{figure=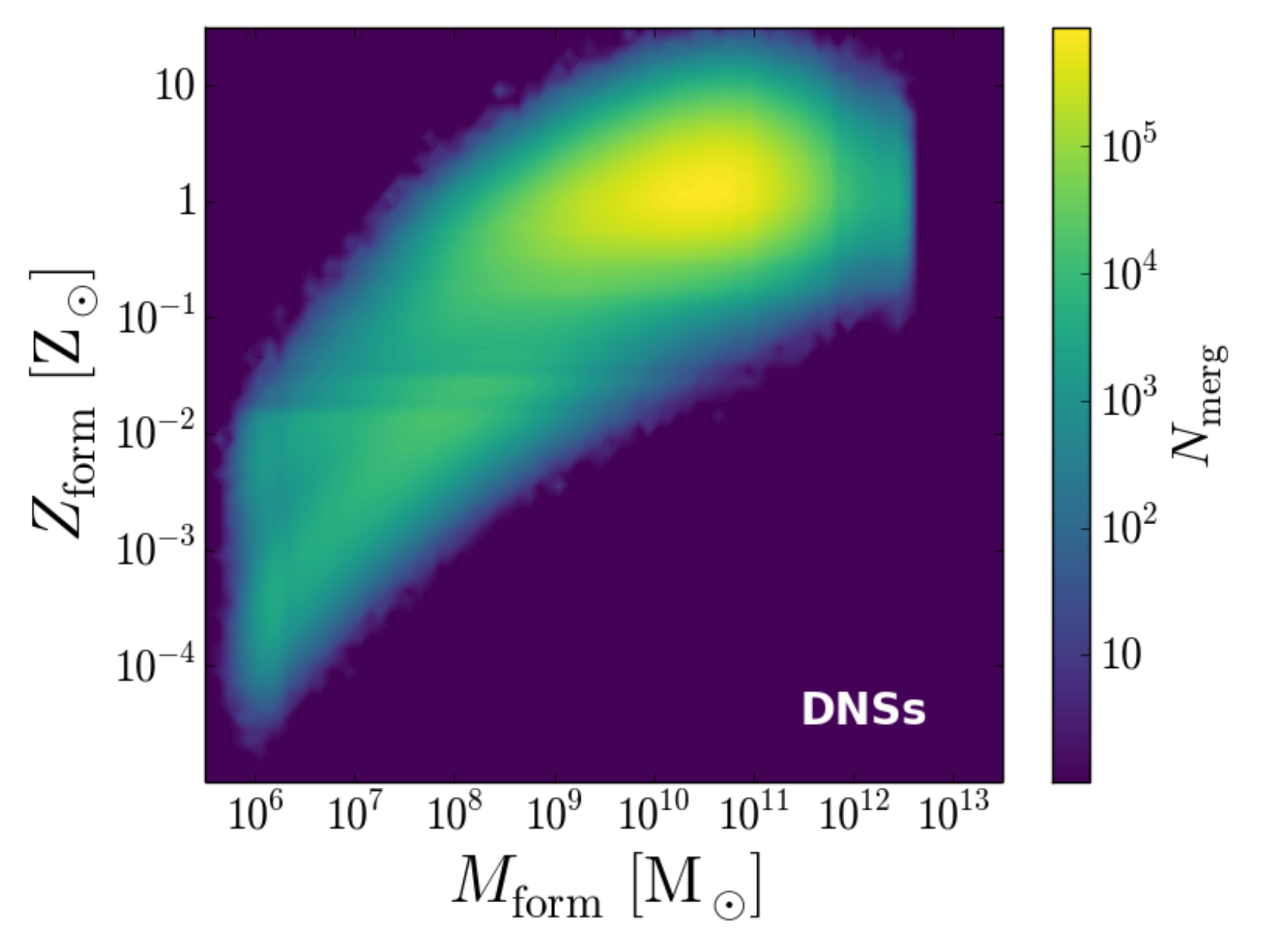,width=9cm} 
\epsfig{figure=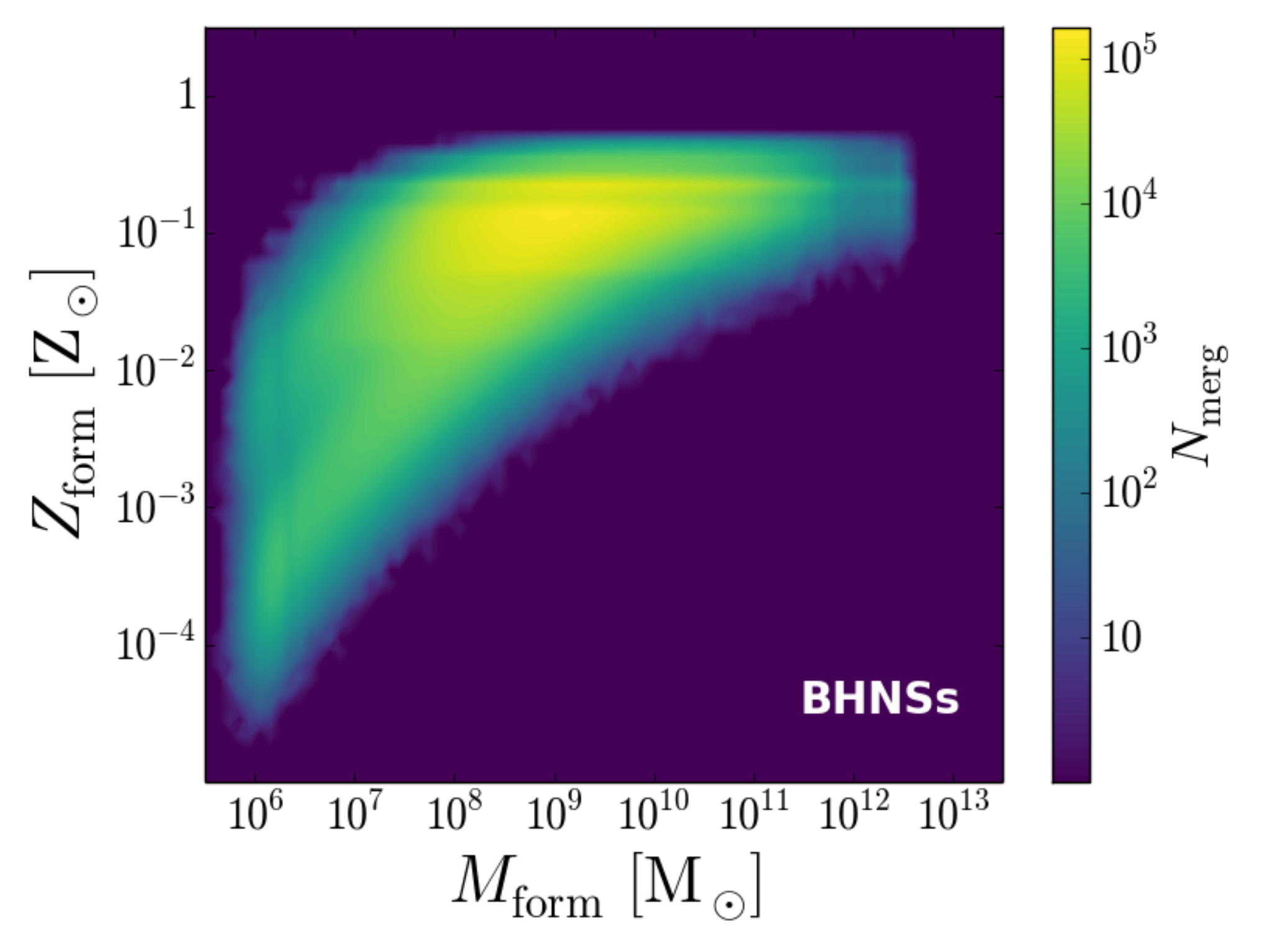,width=9cm} 
\epsfig{figure=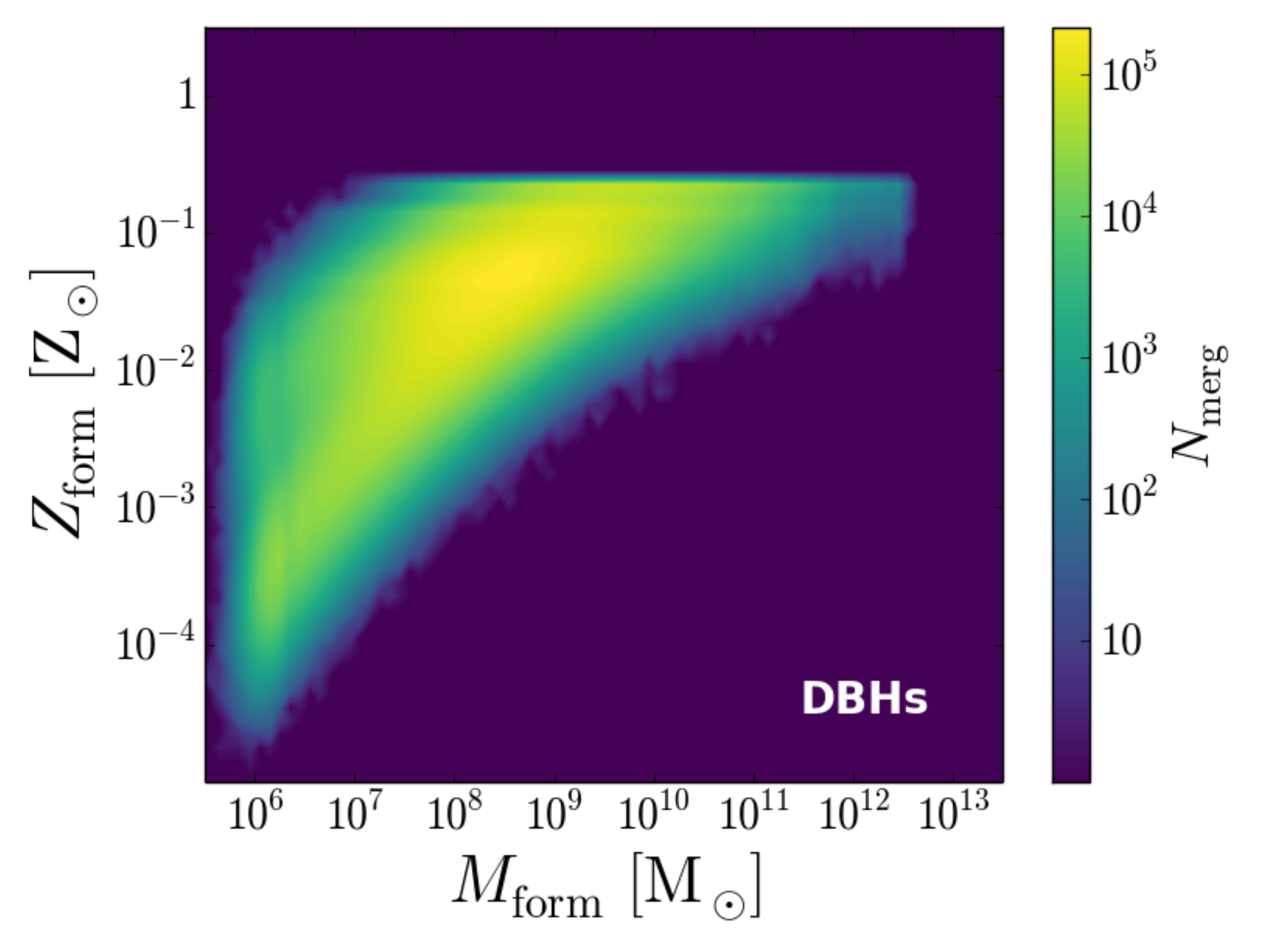,width=9cm} 
}}\caption{\label{fig:figmetal}
Metallicity of the stellar progenitors of merging compact objects ($Z_{\rm form}$) versus $M_{\rm form}$. The colour-coded map (in logarithmic scale) indicates the number of merging compact objects per cell. The cell size is $\log{\delta{}Z_{\rm form}/Z_\odot}\times{}\log{\delta{}M_{\rm form}/M_\odot}=0.1\times{}0.1$. Top panel: merging DNSs; central panel: merging BHNSs; bottom panel: merging DBHs.}  
\end{figure}

This interpretation is confirmed by Figure~\ref{fig:figmetal}, where we show the metallicity of the stellar progenitors of the merging compact objects versus $M_{\rm form}$. These plots are reminiscent of the mass--metallicity relation by construction. The progenitors of merging DNSs form mostly in large galaxies and tend to have solar or super-solar metallicity. In contrast, the progenitors of BHNSs and especially DBHs form mostly at low metallicity ($Z\lesssim{}0.2$ Z$_\odot$) and in intermediate to low-mass galaxies.

\begin{figure}
\center{{
\epsfig{figure=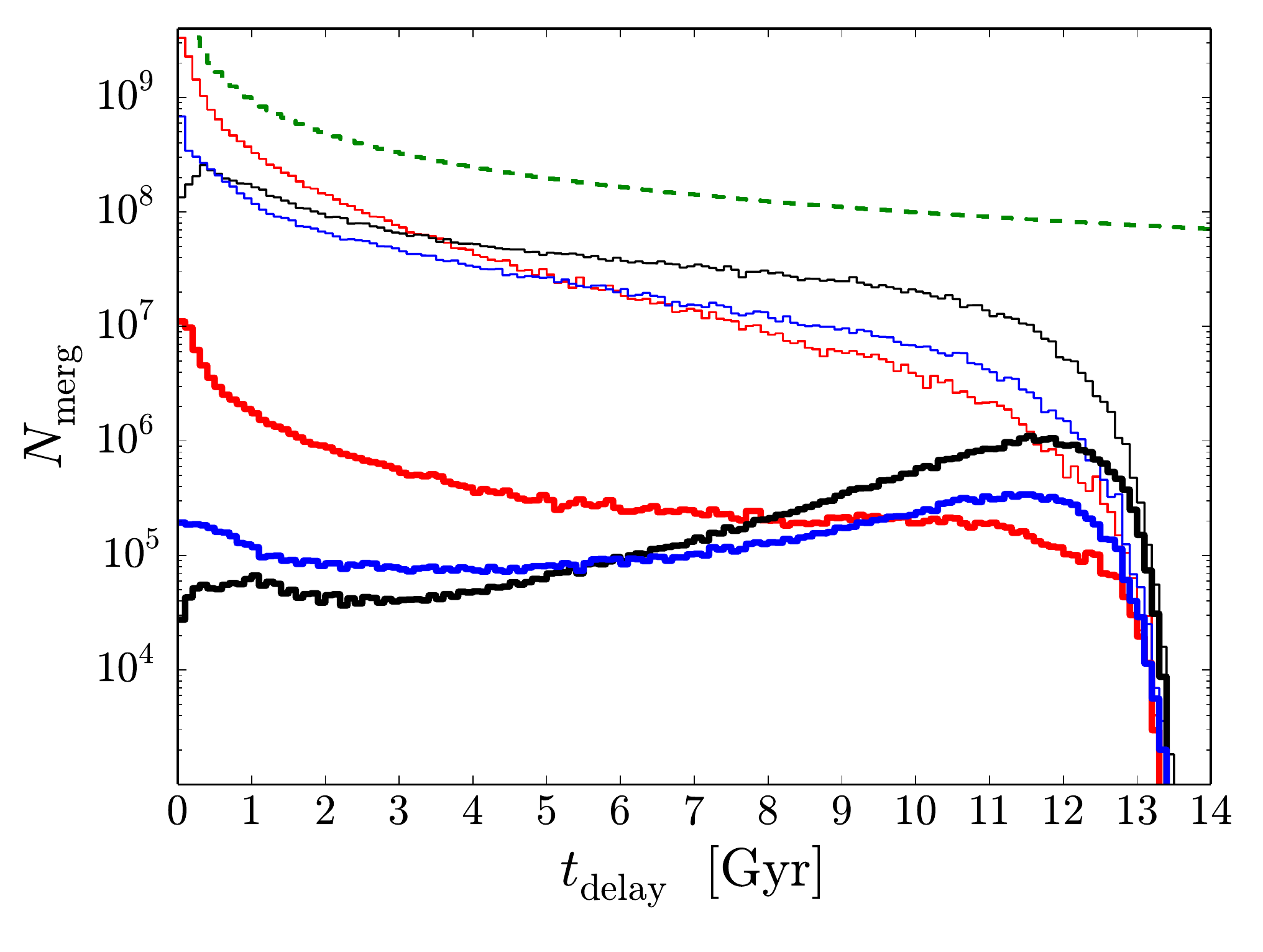,width=9cm} 
}}
\caption{\label{fig:figtime}
Distribution of delay times ($t_{\rm delay}$) of DNSs (red), BHNSs (blue) and DBHs (black). The delay time is defined as the time elapsed from the formation of the stellar progenitors to the merger of the compact objects. Thick solid lines: DNSs, BHNSs and DBHs merging in the local Universe ($z\leq{}0.024$). Thin solid lines: all DNSs, BHNSs and DBHs merging within a Hubble time in the cosmological simulation. Dashed green line: $dN/dt\propto{}t^{-1}$.}
\end{figure}

 Finally, Figure~\ref{fig:figtime} shows the delay time $t_{\rm delay}$ for all DNSs, BHNSs and DBHs merging at $z<0.024$ (thick lines) compared to the delay time of all DNSs, BHNSs and DBHs merging within a Hubble time in our simulation (thin lines).

  We remind that if we consider a coeval binary population, the distribution of delay times of DBH, BHNS and DNS mergers scales approximately as $dN/dt\propto{}t^{-1}$  (see e.g. Figure~6 of \citealt{giacobbo2018b}),  
but figure~\ref{fig:figtime} does not refer to  coeval populations. 

 The distribution of $t_{\rm delay}$ of all DNSs merging in the cosmological simulation (i.e. integrated over all merger redshifts) is even steeper than $t^{-1}$, while the $t_{\rm delay}$ distributions of all DBHs and BHNSs are flatter. On the other hand, all distributions shown as thin lines have a similar trend, a large fraction of merging systems having relatively small delay time ($\lesssim{}1$ Gyr).

In contrast, if we look at the delay times of binaries merging in the local Universe (thick lines), the one of DNSs is very different from those of BHNSs and DBHs. 
The delay time of DNSs merging at low redshift scales approximately as $t^{-1}$: this indicates that most merging DNSs we observe at redshift $z<0.024$ have short delay time ($<3$ Gyr) and merge where they formed.

The distribution of delay time of DBHs merging at $z<0.024$ is bimodal with a main peak at $t_{\rm delay}\sim{}12$ Gyr and a secondary peak at $\sim{}1$ Gyr (Fig.~\ref{fig:figtime}). The secondary peak corresponds to DBHs formed in local metal-poor dwarf galaxies. Thus, the majority of DBHs merging in the local Universe has a long delay time of several Gyr.  This means that most DBHs merging today formed at higher redshift and that the efficiency of DBH formation was higher in the past than today. 

The distribution of delay times of  BHNSs merging at $z<0.024$ is similar to that of DBHs, but less extreme. It is bimodal with a primary peak at $\sim{}12$ Gyr and a secondary peak at $<1$ Gyr. The distribution of delay times confirms the interpretation that most DBHs and BHNSs merging in the local Universe come from high-redshift metal-poor galaxies and only a small fraction forms in local metal-poor dwarf galaxies.

\section{Discussion and conclusions}
We investigated the host galaxies of DNS, BHNSs and DBHs merging in the local Universe, by combining the results of population-synthesis simulations with information from the Illustris-1 cosmological box \citep{vogelsberger2014a,vogelsberger2014b}. To perform the population-synthesis simulations, we have used the {\sc MOBSE} code \citep{giacobbo2018,giacobbo2018b,giacobbo2018c}, which includes up-to-date prescriptions for stellar winds, core-collapse SNe, electron-capture SNe and (pulsational) pair-instability SNe. The masses and rates of merging compact objects simulated with {\rm MOBSE} are consistent with the values inferred from LIGO-Virgo detections \citep{mapelli2017,mapelli2018,giacobbo2018b}.

We find that DNSs merging in the local Universe form preferentially in large galaxies, with stellar mass $10^9-10^{12}$ M$_\odot$ (Fig.~\ref{fig:figmergmform}), and merge mostly in the same galaxy where they formed, with a short delay time (Figs.~\ref{fig:figmergmform_contour} and ~\ref{fig:figtime}). This mass range of the host galaxies is consistent with the mass range of short gamma-ray burst hosts \citep{leibler2010}.

This trend can be explained by the fact that the statistics of DNS mergers is not particularly sensitive to progenitor's metallicity, as shown by \cite{giacobbo2018b}. Thus, DNSs tend to merge more frequently in giant galaxies, where most of the stellar mass is located. Since the delay time scales approximately as $t^{-1}$ (Fig.~\ref{fig:figtime}), most local DNSs merge where they formed. 

In contrast, BHNSs and DBHs merging in the local Universe form preferentially in small galaxies ($<10^{10}$ M$_\odot$) and then merge either in small galaxies or in larger ones (Figs.~\ref{fig:figmergmform_contour} and \ref{fig:figmergmform}).  This result has important implications for the connection between short GRBs and BHNSs. In fact, the mass range of merging BHNS host galaxies is significantly different from the mass range of both short-GRB hosts and merging DNS hosts. This suggests that BHNS mergers cannot explain the bulk of short-GRB population, in agreement with the conclusions of previous papers based on magneto-hydrodynamical relativistic simulations (e.g., \citealt{giacomazzo2013}).

This trend can be explained by the fact that the statistics of DBHs and BHNSs strongly depends on progenitor's metallicity: we expect $\sim{}10^3$ more mergers from a population of metal-poor stars ($Z<0.002$) than from an equivalent population of metal-rich stars ($Z\gtrsim{}0.02$, \citealt{giacobbo2018}). Metal-poor stars form preferentially in local dwarf galaxies and in high-redshift galaxies. Thus, in the local Universe we expect to observe the merger of DBHs and BHNSs formed in high-redshift galaxies with a long delay time, or formed in local dwarf galaxies with a short delay time. Most of the metal-poor high-redshift galaxies have merged into larger galaxies in the local Universe: this explains why a large fraction of DBHs and BHNSs form in small galaxies and reach coalescence in larger ones.

The distribution of delay times of DBHs and BHNSs merging in the local Universe appears to be very different from the $\propto{}t^{-1}$ trend we expect for compact object mergers. This happens because the intrinsic distribution of delay times from population-synthesis simulations scales approximately as $t^{-1}$; but if we restrict our analysis only to mergers happening at low redshift these are dominated by DBHs and BHNSs which formed in the early metal-poor Universe and merge today with a long delay time. In contrast, we do not have this selection effect for DNSs (which are insensitive to metallicity) and thus their delay time distribution scales as  $t^{-1}$ even if we consider only low-redshift mergers.

In our results, the number of merging DBHs and BHNSs starts declining for $M_{\rm form}\lesssim{}10^8$ M$_\odot$. This might be a bias due to the mass resolution of the Illustris. To check whether smaller galaxies can give an even higher contribution to the population of DBHs and BHNSs we will consider higher-resolution smaller-box simulations in future works. 

Another {\it caveat} we should mention is that our current method does not account for binaries which are ejected from their parent galaxy because of natal or dynamical kicks (e.g. \citealt{perna2002,mapelli2011,mapelli2013,giacobbo2018c}). Thus, we cannot investigate whether these events are responsible for the hostless short gamma-ray burst population described by e.g. \cite{fong2013}.

Our findings  are in fair agreement with most previous work (e.g. \citealt{elbert2017,schneider2017,cao2018}) but also suggest that the situation could be more complex than previously thought. In particular, there might be significant differences between the environment of DNSs and that of both DBHs  and BHNSs. Overall, the metallicity of progenitor stars is a key property for DBHs and BHNSs, while it is much less important for DNSs.


Our results have strong implications for forthcoming detections of compact-object mergers: DNS mergers in the local Universe should happen mostly in galaxies with stellar mass $\sim{}10^{9}-10^{12}$ M$_\odot$ (with a sharp peak at $\sim{}10^{10}-10^{11}$ M$_\odot$), while BHNSs and DBHs are expected to coalesce in galaxies with a broad range of masses from $\sim{}10^8$ M$_\odot$ up to few $\times{}10^{12}$  M$_\odot$. Further analysis should assess whether these galaxies are mostly early or late type, field galaxies or group/cluster galaxies.

Our final goal is to constrain the environment of compact-object mergers and to provide clues on the host galaxies of GW events. In particular, our results could be combined with a low-latency localization algorithm (e.g.  the Dirichlet process Gaussian-mixture model described in \citealt{delpozzo2018}) to produce a list of most probable host galaxies of a new GW detection, facilitating a prompt multimessenger follow-up (e.g. \citealt{nissanke2013,hanna2014,gehrels2016,singer2016}). Alternatively, if the electromagnetic counterpart of a GW event is not observed, our results could be used to ``weight'' possible host galaxies within the LIGO-Virgo error box, providing useful information for measurements of the Hubble constant (e.g. \citealt{schutz1986,delpozzo2012,chen2017,delpozzo2018}).


\section*{Acknowledgments}
We thank Christopher Berry, Giulia Stratta, Irina Dvorkin and Elena D'Onghia for useful discussions. We also thank the anonymous referee for their useful comments. We warmly thank The Illustris team for making their simulations publicly available. Numerical calculations have been performed through a CINECA-INFN agreement and through a CINECA-INAF agreement, providing access to resources on GALILEO and MARCONI at CINECA.
 MM  acknowledges financial support from the MERAC Foundation through grant `The physics of gas and protoplanetary discs in the Galactic centre', from INAF through PRIN-SKA `Opening a new era in pulsars and compact objects science with MeerKat', from MIUR through Progetto Premiale 'FIGARO' (Fostering Italian Leadership in the Field of Gravitational Wave Astrophysics) and 'MITiC' (MIning The Cosmos  Big Data and Innovative Italian Technology for Frontier Astrophysics and Cosmology), and from the Austrian National Science Foundation through FWF stand-alone grant P31154-N27 `Unraveling merging neutron stars and black hole - neutron star binaries with population-synthesis simulations'. NG acknowledges financial support from Fondazione Ing. Aldo Gini and thanks the Institute for Astrophysics and Particle Physics of the University of Innsbruck for hosting him during the preparation of this paper.
 This work benefited from support by the International Space Science Institute (ISSI), Bern, Switzerland,  through its International Team programme ref. no. 393
 {\it The Evolution of Rich Stellar Populations \& BH Binaries} (2017-18).
 
\bibliography{./bibliography}

\end{document}